\documentstyle[aps,preprint,epsfig]{revtex}
\title{Component separation in harmonically trapped boson-fermion mixtures}
\author{Nicolai Nygaard and Klaus M{\o}lmer\\
{\small Institute of Physics and Astronomy, University of Aarhus }\\
{\small DK-8000 {\AA}rhus C, Denmark}}
\begin{document}
\maketitle

\begin{abstract}
We present a numerical study of mixed boson-fermion systems at zero temperature
in isotropic and anisotropic harmonic traps. We investigate the phenomenon of 
component separation as function of the strength of the inter-particle interaction.
While solving a Gross-Pitaevskii mean field equation for the boson distribution 
in the trap, we utilize two different
methods to extract the density profile of the fermion component; 
a semiclassical Thomas-Fermi approximation and a quantum mechanical Slater 
determinant Schr\"{o}dinger equation.
\end{abstract}


\narrowtext
\tighten

\section{Introduction}
Since the recent experimental realization of Bose-Einstein condesation in dilute
gases of rubidium~\cite{cornell2,heinzen,konstanz,quictrap}, sodium~\cite{ketterle1,hau2}, 
lithium~\cite{hulet1}, and hydrogen~\cite{hydrogen}
a great deal of interest in Bose condensed systems has concentrated on the topic
of multi-component condensates. This field was stimulated by the succesful demonstration 
of overlapping condensates in different spin states of rubidium in a magnetic
trap~\cite{cornell3,cornell5} and of sodium in an optical trap~\cite{ketterle8}, the
(binary) mixtures being produced either by sympathetic cooling, which involves one species
being cooled to below the transition temperature only through thermal contact
with an already condensed Bose gas, or by radiative transitions out of a single 
component condensate. Since then a host of experiments has been conducted
on systems with two condensates, exploring both the dynamics of component
separation~\cite{cornell6}, and measuring the relative quantum phase of the two 
Bose-Einstein condensates~\cite{cornell7}.
Most of the theoretical work concerning multi-component 
condensates
~\cite{shenoy,esry,walls2,bigelow2,sten2,bigelow3,bigelow4,cornell10,gordon,shlyapnikov4}
has been devoted to systems of two Bose condensates. 
However, other systems are of fundamental interest,
one of these being a Bose condensate with fermionic impurities, a system reminiscent of
superfluid \mbox{$^{3}$He-$^{4}$He} mixtures.
In particular the possibilty of sympathetic cooling of fermionic isotopes
has been predicted in both $^6\mathrm{Li}$-$^7\mathrm{Li}$~\cite{lithium},
$^{39}\mathrm{K}$-$^{40}\mathrm{K}$, and $^{41}\mathrm{K}$-$^{40}\mathrm{K}$~\cite{potassium}.
Magneto-optical trapping of the fermionic potassium isotope $^{40}\mathrm{K}$ has been 
reported~\cite{cornell12}.

The boson-fermion mixture was discussed in a
previous paper~\cite{klaus1} within the Thomas-Fermi approximation, which
amounts to neglecting the kinetic energy of the bosons, and to apply a semi-classical
filling of phase space of the fermions. For the bosons, this is a valid approximation 
in the limit of strong interactions or large particle numbers, see~\cite{lewenstein3}.
In this paper we present a numerical analysis of the system, incorporating the
correct operator form of the kinetic energy of the particles.

The paper is structured as follows. In Sec. II we study in detail the case of an
isotropic external potential and we develop both the Thomas-Fermi approximation and the
full quantum mechanical description of the fermions. The numerical procedure
is briefly introduced. In Sec. III the case
of the anisotropic harmonic oscillator trap is outlined within 
the Thomas-Fermi approximation for the fermions. In Sec. IV we
present our quantitative results for the isotropic and anisotropic trapping potentials, 
demonstrating the
accuracy of the predictions made in~\cite{klaus1}, and addressing the issue 
of symmetry breaking in elongated traps. Sec. V summarizes the
main results.

Throughout, we assume that the bosons and fermions have the same mass, $M$, and 
that the atoms are all trapped in the same external harmonic oscillator potential.
This choise is of course only a convenience; all our calculations are readily generalized
to differing experimental parameters.

\section{Isotropic traps}

\subsection{Gross-Pitaevskii equation for the bosons}
In the mean field description the behavior of the single particle wavefunction
$\psi(\vec{r})$, assumed to describe all $N_{B}$ bosons in the gas, is governed by the 
Gross-Pitaevskii equation~\cite{pitaevskii,gross}. In the presence of
fermions, this equation is modified by the 
addition of an interaction term proportional to the fermion density, $n_{F}(\vec{r})$
\begin{equation}
\left[ -\frac{\hbar^2}{2M}\nabla^2+V_{ext}(\vec{r})+gN_{B}|\psi(\vec{r})|^2
+hn_{F}(\vec{r})\right]\psi(\vec{r}) = \mu \psi(\vec{r}),
\label{gross-pita}
\end{equation}
where $V_{ext}(\vec r)
={1\over 2}M(\omega_x^2 x^2 + \omega_y^2y^2 + \omega_z^2 z^2)$
is the external confining potential, and $\mu$ is the boson chemical
potential (energy per particle). The value of $\mu$ is fixed by the 
normalisation condition, $\int d^{3}r \, n_{B}(\vec{r}) = N_{B}$ on the boson
density $n_{B}(\vec{r}) = N_{B}|\psi(\vec{r})|^2$. 

The low kinetic energies of the atoms
permit the replacement of their short range interaction potential by
a delta function potential of strength g or h. This is known as the 
pseudopotential method~\cite{huang}. There is no fermion-fermion
interaction in this description, see below.
In (\ref{gross-pita}) \mbox{g and h} thus represent
the boson-boson and the boson-fermion interaction strengths
proportional to the respective $s$-wave scattering lengths~\cite{lewenstein3}.

In isotropic traps we have $V_{ext}(\vec{r}) = {1\over 2}M\omega^2r^2$, r being the
distance from the trap center. By the substitution $\chi = r\psi$ in (\ref{gross-pita})
we obtain the radial equation
\begin{equation}
-\frac{\hbar^2}{2M}\frac{d^2\chi}{dr^2}+\left[V_{ext}(r)+gN_{B}
\left|\frac{\chi(r)}{r}\right|^2
+hn_{F}(r)\right]\chi(r) = \mu \chi(r).
\label{rad-bos}
\end{equation}
In order to simplify the formalism, we rescale (\ref{rad-bos})
in terms of harmonic oscillator units, that is
\begin{eqnarray}
\vec{r} &  = & a_0\tilde{\vec{r}}, \nonumber \\
\mu  & = & \hbar\omega\tilde{\mu}, \nonumber \\
\psi(\vec{r}) & = & a_0^{-3/2}\tilde{\psi}(\tilde{\vec{r}}), \nonumber \\
\chi(r) & = & a_0^{-2}\tilde{\chi}(\tilde{r}),
\label{ho-units}
\end{eqnarray}
where $a_0 = \sqrt{\hbar/M\omega}$ is the width of the oscillator ground state.
Defining
\begin{equation}
\tilde{g} = \frac{gM}{a_0\hbar^2} , \ \tilde{h} = \frac{hM}{a_0\hbar^2},
\label{intst-tilde}
\end{equation}
we arrive at the simplified equation for the radial function
\begin{equation}
-{1\over 2}\frac{d^2\tilde{\chi}}{d\tilde{r}^2}+\left[{1\over 2}\tilde{r}^2
+\tilde{g}N_{B}\left|\frac{\tilde{\chi}(\tilde{r})}{\tilde{r}}\right|^2
+\tilde{h}\tilde{n_{F}}(\tilde{r})\right]\tilde{\chi}(\tilde{r}) 
= \tilde{\mu} \tilde{\chi}(\tilde{r}).
\label{rad-tilde}
\end{equation}
In the remaining parts of the paper we shall omit the tilde from this equation.
 
To solve the Gross-Pitaevskii equation for the bosons, we
must find $n_{F}(\vec{r})$. To this end we invoke two methods: A semi-classical
(Thomas-Fermi) approximation and a quantum mechanical treatment. 

\subsection{Thomas-Fermi approximation for the fermions}
In the semi-classical (Thomas-Fermi) approximation the particles are
assigned classical positions and momenta, but the effects of quantum
statistics are taken into account. That is: The density in the occupied part of
phase space is
simply $(2\pi)^{-3}$, and sums over states can be replaced by the corresponding 
integrals over $\vec{r}$ or $\vec{k}$. 
The fermions experience a potential $V(\vec r)=V_{ext}(\vec r)+hn_B(\vec r)$ and
for particle motion in such a potential it is posible to
define a local Fermi vector $\vec{k}_{F}(\vec{r})$ by
\begin{equation}
E_F={\hbar^2k_F(\vec r)^2\over 2M}+V(\vec r),
\label{energy}
\end{equation}
so that the volume of the
local Fermi sea in $k$ space is simply
\begin{equation}
{4\over 3}\pi k_F(\vec r)^3=(2\pi)^3 n_F(\vec r).
\label{fermi-k}
\end{equation} 
In the low temperature limit, where $p$-wave (and higher multipole) scattering
can be neglected, the supression of the $s$-wave scattering amplitude
due to the antisymmetry of the many-body wavefunction implies that the spin polarized
fermions constitute a noninteracting gas (for the case of an 
interacting Fermi gas, see~\cite{burnett6}). Hence the
density of the fermionic component is given by
\begin{equation}
n_F(\vec r)=\left\{{2M\over \hbar^2}\left[E_F-V_{ext}(\vec r)-hn_{B}(\vec r)\right]
\right\}^{3/2}/(6\pi^2).
\label{density-f}
\end{equation}
As in the case of the bosons, where the chemical potential must be adjusted
for the integral of the density over space to give the correct number of particles,
the Fermi energy determines the proper normalisation; 
$\int d^{3}r \, n_{F}(\vec{r}) = N_{F}$.
For a thorough discussion of trapped fermions (also at $T > 0$), and
comments on the range of validity of the Thomas-Fermi approximation see~\cite{fermigas}.

\subsection{Slater determinant description}
The many-body wavefunction, $\Psi(\vec{r}_1 \ldots \vec{r}_{{N}_F})$, may be represented
by a Slater determinant
\begin{equation}
\Psi(\vec{r}_1 \ldots \vec{r}_{{N}_F}) = \frac{1}{\sqrt{N!}}{\mathcal{A}} \prod_{i=1}^{N_F}
\varphi_i(\vec{r}_i),
\label{slater}
\end{equation}
where ${\mathcal{A}}$ is the antisymmetrization operator.
This Slater determinant solves a stationary Schr\"{o}dinger equation
\begin{equation}
\hat{H}\Psi(\vec{r}) = E\Psi(\vec{r}),
\label{stat-schr}
\end{equation}
with a Hamiltonian that is the sum of $N_F$ independent single-particle operators
\begin{eqnarray}
\hat{H} &  = & \sum_{i=1}^{N_F} \hat{H}_i, \\ 
\hat{H}_i &  = & -\frac{\hbar^2}{2M}\nabla_{r_i}^2+{1\over 2}M\omega^2r_{i}^2
+hn_{B}(\vec{r}_i).
\label{hamilton}
\end{eqnarray}
The orbitals $\varphi_i(\vec{r}_i)$ solve the eigenvalue equation
\begin{equation}
\hat{H}_i\varphi_i(\vec{r}_i) = E_i\varphi_i(\vec{r}_i).
\label{eigeneq}
\end{equation}
We make the substitution $\varphi(\vec{r}) = 
\frac{u_{n\ell}(r)}{r}Y_{{\ell}m}(\theta,\phi)$, where $Y_{lm}(\theta,\phi)$
are the usual spherical harmonics, and we thus obtain a radial equation
for the functions $u_{n\ell}$ in harmonic oscillator units
\begin{equation}
-{1\over 2}\frac{d^2u_{n\ell}}{dr^2}+\left[{1\over 2}r^2+\frac{\ell(\ell+1)}{2r^2}
+hn_{B}(r)\right]u_{n\ell}(r) = E_{n\ell} \, u_{n\ell}(r).
\label{rad-fer}
\end{equation} 
It is important to keep in mind that  the radial functions must satisfy the 
boundary condition $u_{n\ell}(0) = 0$, to ensure a finite particle density
at the center of the trap.

Equation~(\ref{rad-fer}) can be solved once for every $\ell$-value, thus producing the 
energy spectrum. The centrifugal term in the radial equation, implies that the fermions
can be considered to move in an isotropic effective potential, 
$V_{eff}(r) = V_{ext}(r)+hn_B(r)+\frac{\ell(\ell+1)}{2r^2}$.  
The energy levels $E_{n\ell}$ are
$2\ell+1$ times degenerate, and we have the fermion density given by
\begin{equation}
n_F(\vec{r}) = \sum_{\stackrel{occupied}{states}} \left| \frac{u_{n\ell}(r)}{r}
Y_{{\ell}m}(\theta,\phi)\right|^2 = \sum_{n\ell} (2\ell +1)
\frac{|u_{n\ell}(r)|^2}{4\pi r^2},
\label{fermidens}
\end{equation}
since $\sum_{m = -\ell}^{m = \ell} |Y_{{\ell}m}(\theta,\phi)|^2 = (2\ell+1)/4\pi$.
Once found the eigenstates are sorted after energy and the energy levels are filled
from below with $N_F$ particles.
The Fermi energy is the energy of the highest occupied orbital.
The maximum value of the angular momentum, may be estimated from the 
Thomas-Fermi expressions (\ref{fermi-k},\ref{density-f}) for $h=0$ by 
maximizing $r k_F (\vec{r})$, the maximal length of $\ell$ at the point $\vec r$.
This yields the simple result, ${\ell}_{max} \! \approx \! E_F$, where the Fermi energy
in the noninteracting limit is $E_F=(6N_F)^{1/3}$ in harmonic oscillator units.
To test our numerical calculations for fermions not interacting with the bosons 
($h=0$), we have compared our spatial density distributions with those of 
Schneider and Wallis~\cite{meso_fermi} and found excellent agreement.

\subsection{Numerical Procedure}
We note that the solution of both Eqs.~(\ref{density-f}) and~(\ref{rad-fer})
require prior knowledge of the boson density $n_B(\vec r)$.
To obtain the density profiles of the two components,
we insert iteratively the density of one component into the equation for the other
until a desired convergence is reached.

To solve~(\ref{rad-tilde}) for the boson density,
we use the method of steepest descend, that is we propagate a trial function (which can
be chosen initially almost arbitrarily) in imaginary time
$\tau$, replacing $\mu\chi(r)$ by $-(\partial / \partial\tau)\chi(r,\tau)$. 
In the long time limit the propagation ``filters'' the trial function 
to the condensate ground state
Alternative methods for solving numerically the \mbox{Gross-Pitaevskii} equation are 
presented in~\cite{bigelow3,burnett7,burnett1,krauth}.

The evaluation of the fermion density profile is done by two methods, as described above.
In the case of the Thomas-Fermi approximation, $n_F(\vec r)$ is found by direct insertion
of $n_B(\vec r)$ into~(\ref{density-f}), searching numerically for the energy $E_F$
giving the right number of particles. Within the Slater determinant
method one obtains the density profile directly from~(\ref{fermidens}), once the 
diagonalization of~(\ref{rad-fer}) has been done. 

\section{Anisotropic traps}
In this section we treat the case of an anisotropic trapping potential with 
a cylindrical geometry ($\omega_x \!=\omega_y = \! \omega_{\perp} \neq
\omega_z$) as this corresponds to current experimental 
setups
We thus have
\begin{equation}
V_{ext} = {1\over 2}M\omega_{\perp}^2r^2+{1\over 2}M\omega_z^2z^2,
\label{aniso-pot}
\end{equation}
where $r \! =  \! \sqrt{x^2+y^2}$ and $z$ are the radial and axial
coordinate respectively. We define the asymmetry parameter 
$\lambda = \omega_z / \omega_{\perp}$.

As in the case of the isotropic potential we have for the bosons a 
non-linear Schr\"{o}dinger equation corresponding to~(\ref{gross-pita}). 
By the substitution $\chi = \sqrt{r}\psi$ we obtain the equation
\begin{eqnarray}
\mu\chi(r,z)=&-& {1\over 2} \left[ \frac{\partial^2\chi}{\partial r^2} + \frac{\partial^2\chi}
{\partial z^2}\right]-\frac{\chi(r,z)}{8r^2}
+{1\over 2}(r^2+\lambda^2z^2)\chi(r,z) \nonumber \\
&+& gN_B\frac{\left|\chi(r)\right|^2}{r}\chi(r,z)+hn_F(\vec{r})\chi(r,z),
\label{aniso-gp}
\end{eqnarray}
in harmonic oscillator units. 
Again the radial function has to vanish on the symmetry axis to remove potential 
problems of divergences near the origin. This boundary
condition is implemented in our numerical procedure by imposing on the radial
function a $\sqrt{r}$ dependence for small values of $r$, fitting to the value
of $\chi$ at larger distances from the axis. As in the case of the isotropic trapping
potential a Split-Step-Fourier technique is used to propagate the boson
wavefunction to the ground state. An alternative method for solving
the Gross-Pitaevskii equation in a cylindrical configuration,
applying an alternating-direction
implicit method to compute the derivatives is discussed in~\cite{cooper}. 

We shall limit ourselves to the Thomas-Fermi approximation for the fermions, both out
of necessity and convenience. Already in the spherically symmetric, effectively
1-dimensional case, the full quantum
mechanical analysis is very time consuming and as we shall
demonstrate in the next section, the Thomas-Fermi approximation offers the same
qualitative features as the exact description.
The fermion density is thus evaluated using equation~(\ref{density-f}) with the
external potential~(\ref{aniso-pot}) and with the boson density obtained 
from~(\ref{aniso-gp}).

\section{Results}
The main conclusion of~\cite{klaus1} is the prediction of a
component separation under variation of the strength
of the boson-boson and boson-fermion interaction. In the Thomas-Fermi
approximation for both components, the density distributions solve the
coupled equations
\begin{eqnarray}
V_{ext}(\vec r)+g\cdot n_B(\vec r) + h\cdot n_F(\vec r)&=&\mu\nonumber \\
{\hbar^2\over 2M}(6\pi^2n_F(\vec r))^{2/3}+
V_{ext}(\vec r)+h\cdot n_B(\vec r)& =& E_F.
\label{coupled}
\end{eqnarray}
In the case of $N_F/N_B \ll 1$ the fermions may be neglected in the equation
for the bosons. For the fermions we then obtain the simple equation
\begin{eqnarray}
{\hbar^2\over 2M}(6\pi^2n_F(\vec r))^{2/3}+
(1-{h\over g}) V_{ext}(\vec r) + {h\over g}\mu = E_F,
\label{fewferm}
\end{eqnarray}
where the terms proportional with $h$ are absent in regions with vanishing
$n_B(\vec r)$. We may distinguish between 3 different types of solutions;
if $h < g$ the potential minimum of the fermions is located at the center
of the trap, and if their number is small enough, they will constitute
a 'core' entirely enclosed within the Bose condensate. The two quantum
gases are truly interpenetrating.
If $h = g$ the fermions have a constant density throughout the Bose 
condensate, falling towards zero outside. 
If $h > g$ the effective potential for the fermions
is that of an inverted harmonic oscillator having a minimum 
at the edge of the Bose condensate, where the fermions localize  
as a 'shell' wrapped around the condensate.

\subsection{Isotropic trap, quantum treatment}

\begin{figure}
\centerline{
\epsfig{file=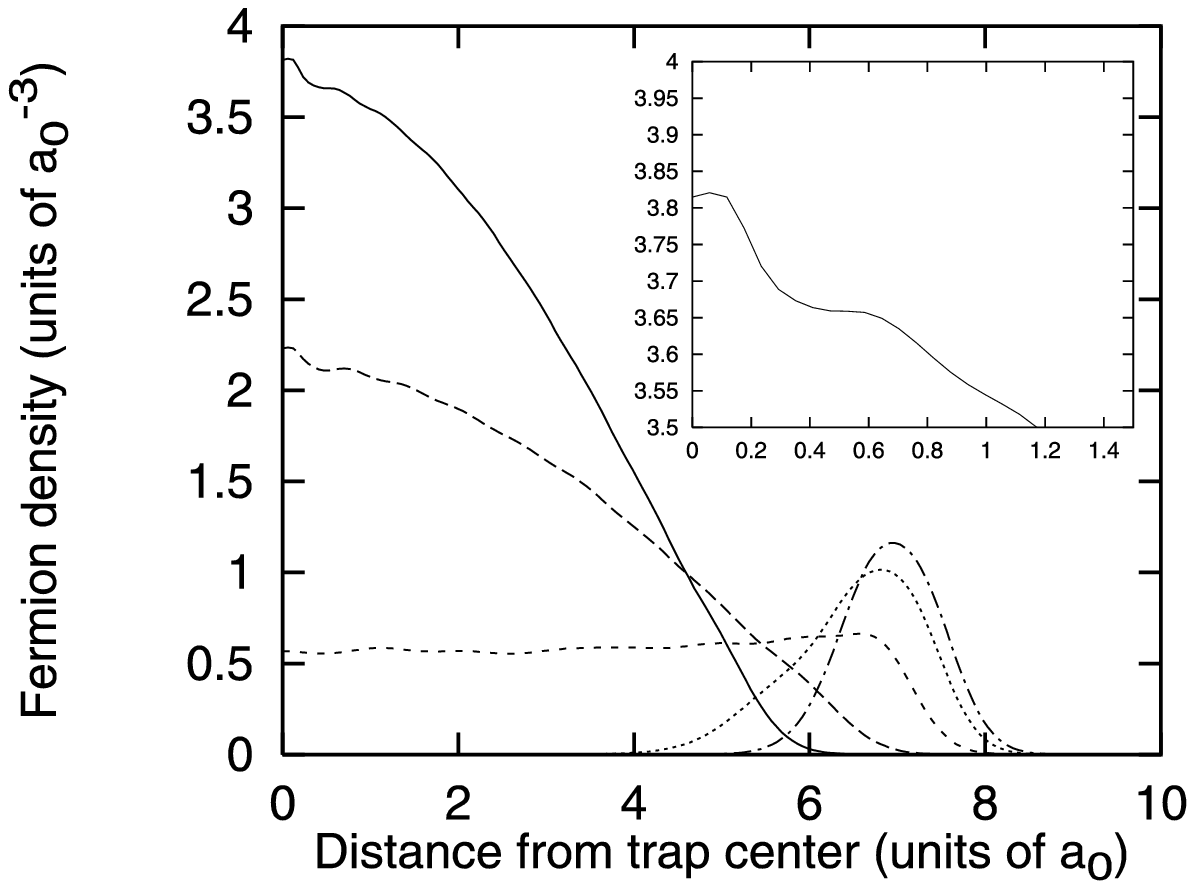,angle=270,width=8.5cm}}
\vspace{5 mm}
\caption{Spatial distribution of 1000 fermions
as function of distance from the trap center. 
$g=0.015 \hbar{\omega}a_0^3$, and the boson-fermion
coupling takes the values, $h=0$  (solid curve), $h=g/2$ (long-dashed curve),
$h=g$ (dashed curve), $h=3g/2$ (dotted curve), and $h=2g$ (dot-dashed curve).
A magnification of part of the top curve is shown in the inset.
The Bose-Einstein condensate component of $10^6$ atoms 
is unaffected by the fermions
and extends out to the distance $\sim 7a_0$, where cusps in the fermion
distributions are visible.}
\label{fermiplot}
\end{figure}

When we replace the Thomas-Fermi approximation by an exact description
including the kinetic energy operator for the bosons and  treating
the fermions quantum mechanically, we expect to observe the same overall
behaviour, but with minor corrections. The boson kinetic energy
is expected to cause penetration into the fermionic component
and a rounding off of the atomic distributions at the boundaries.
Fig.~\ref{fermiplot} shows the spatial distribution of 1000 fermions in a
condensate of $10^6$ bosons for different values of the boson-fermion
interactions strength, $h$. The strength of the boson-boson interaction,
$g$, is chosen to give maximal overlap between the two atomic clouds. In 
order to have clouds of comparable size, we equate the Thomas-Fermi expressions
for the radius of the Bose condensate $(15N_Bg/4\pi M\omega^2)^{1/5}$, and
the radius of the zero temperature Fermi gas
$(48N_F)^{1/6}\sqrt{\hbar/M\omega}$. This gives the condition:
\begin{equation}
g/(\hbar\omega {a_0}^3) \simeq 21.1 {N_F}^{5/6}/N_B, 
\label{g_constraint}
\end{equation}
which for the parameters of Fig.~\ref{fermiplot} requires
$g~=~0.015$. The coupling $g$ differs for different atomic species and
this value is in approximate agreement 
with the coupling stregth in the MIT Na setup~\cite{ketterle1}, and we recall
the possibility to achieve couplings of
arbitrary strength by the recently demonstrated modification
of the atomic scattering length by external 
fields~\cite{shlyapnikov2,ketterle9}. This allows a 'tuning' of 
the scattering length through both positive and negative values.
Finally we recall that we have insisted on equal masses and trapping
potentials for the two components. If these constraints are 
relaxed, we may more easily vary 
the values of the scaled interaction strengths.

The oscillations in the fermion density distribution near the trap center
reflect the matter wave modulation of the particles in the 
outermost shell. Their de Broglie wavelength can be estimated
in the Thomas-Fermi approximation from (\ref{energy}): In the center
of the trap the particles in the $\ell \! = \! 0$ states experience a vanishing
potential for $h=0$. As the Thomas-Fermi expression for the Fermi energy
of $N_F$ fermions in a harmonic potential is $E_F = (6N_F)^{1/3}\hbar\omega$ we find
for the de Broglie wavelength
\begin{equation}
\lambda_{DB} \sim \frac{2\pi}{k_F(0)} \sim \frac{\sqrt{2}\pi a_0}{(6N_F)^{1/6}}
\sim 1 \ a_0,
\end{equation}
an estimate that is reproduced by the data, see inset.

\begin{figure}
$ \begin{array}{ccc}
\epsfig{file=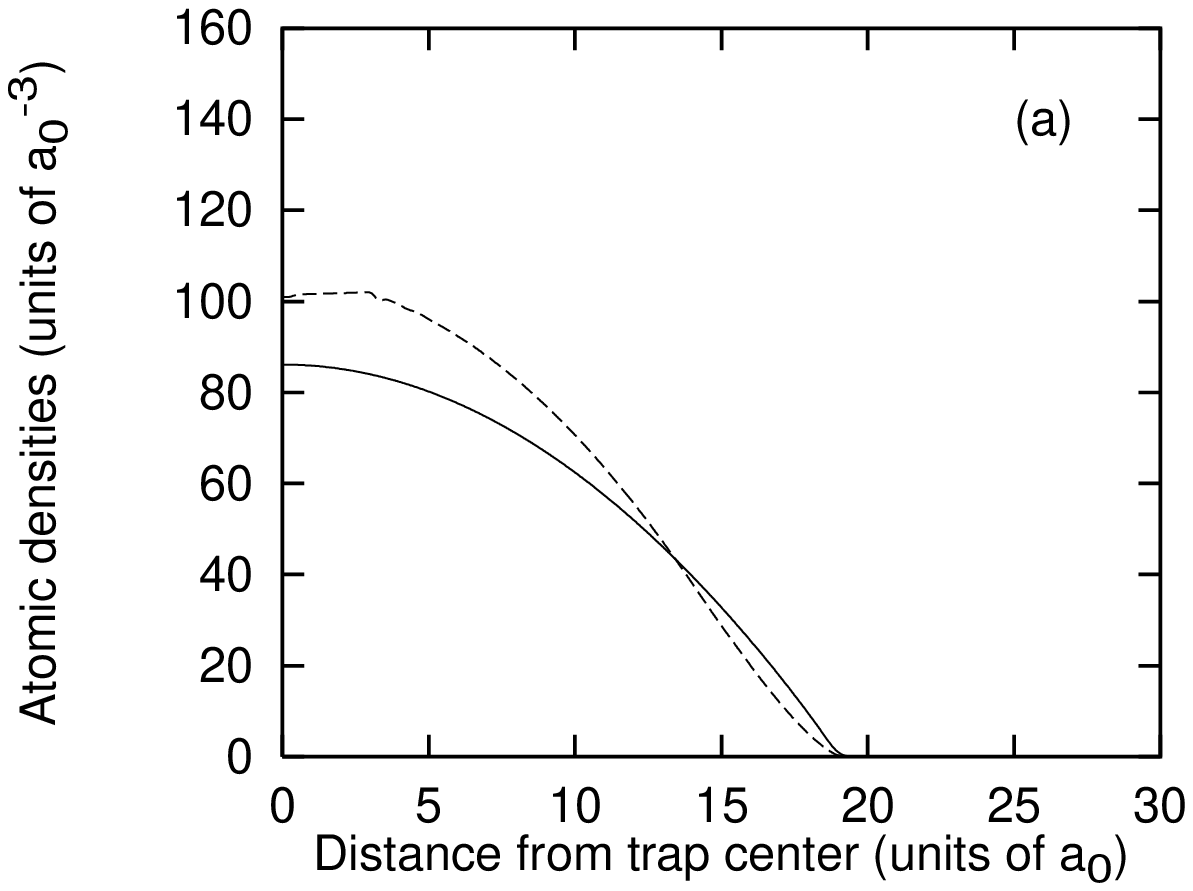,angle=270,width=5.0cm} &
\epsfig{file=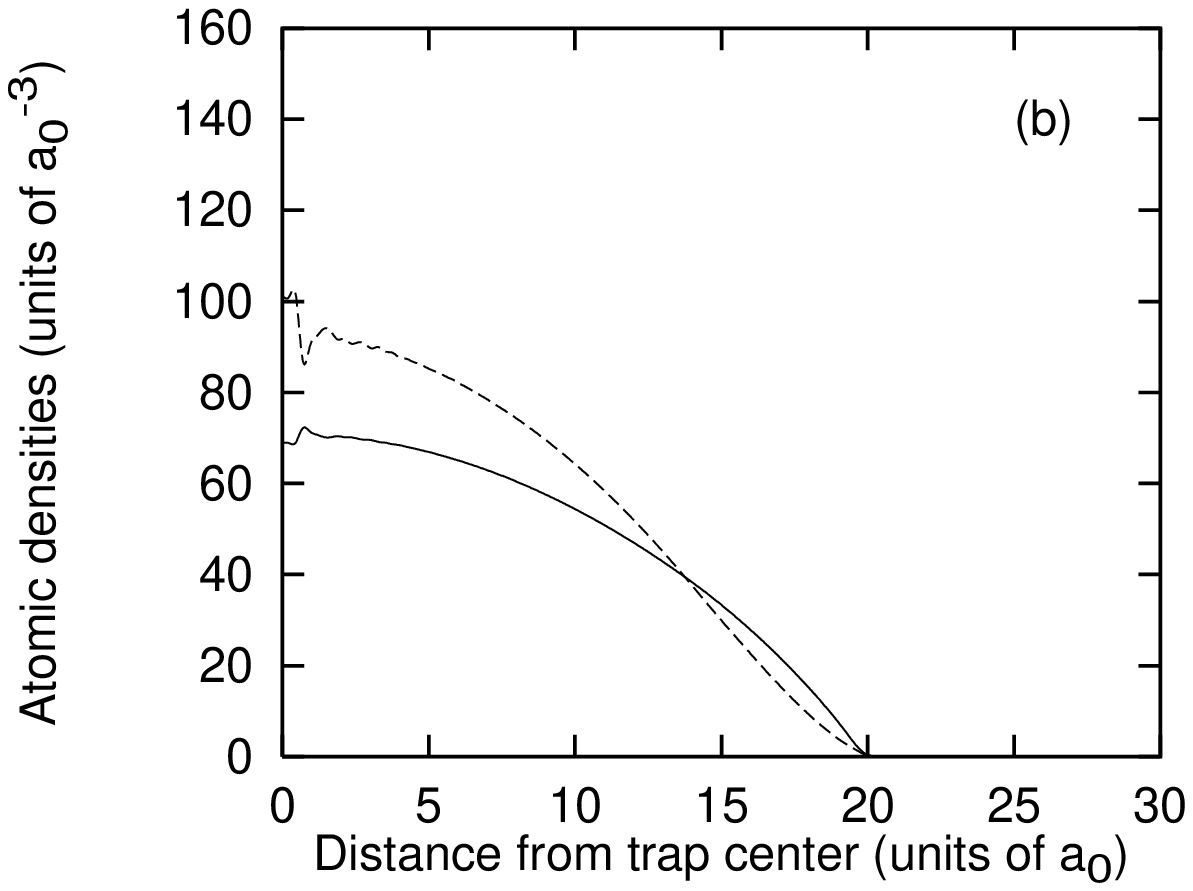,angle=270,width=5.0cm} &
\epsfig{file=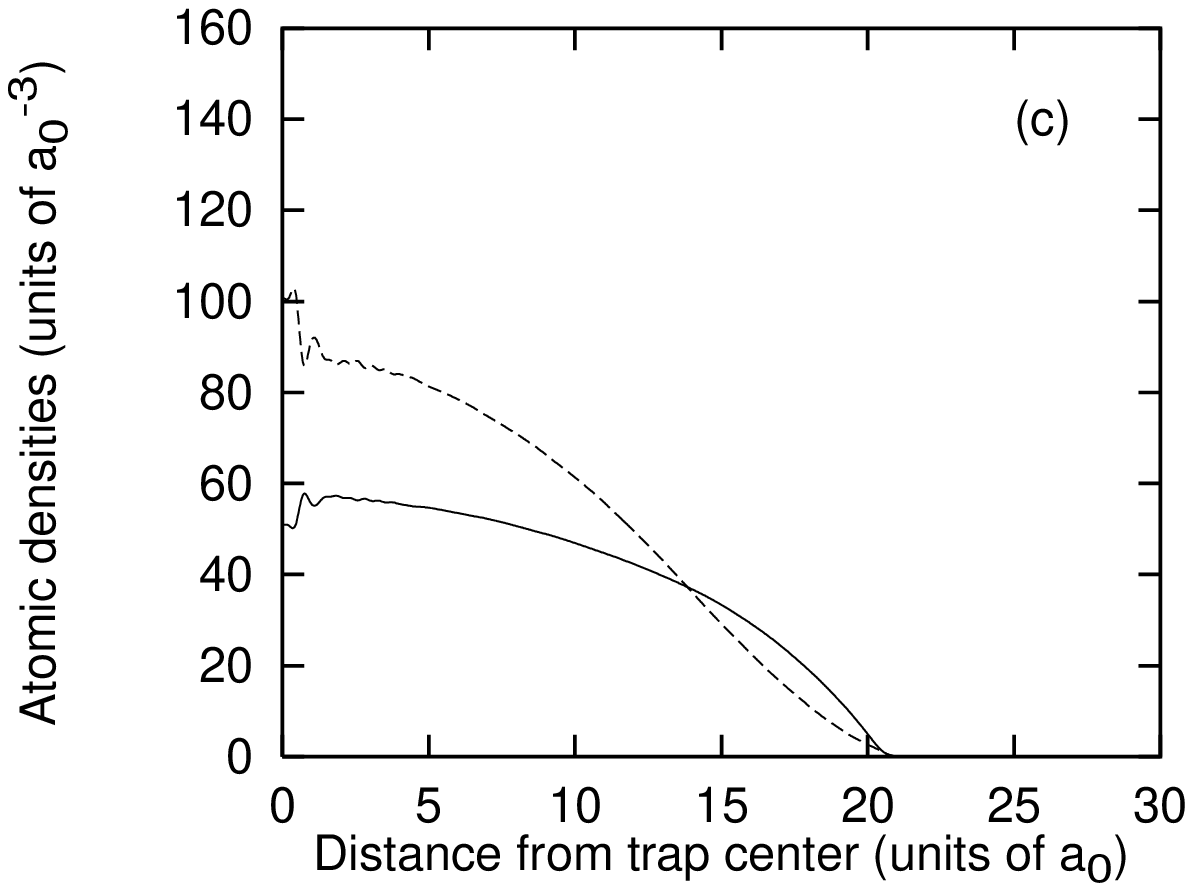,angle=270,width=5.0cm} \\
\epsfig{file=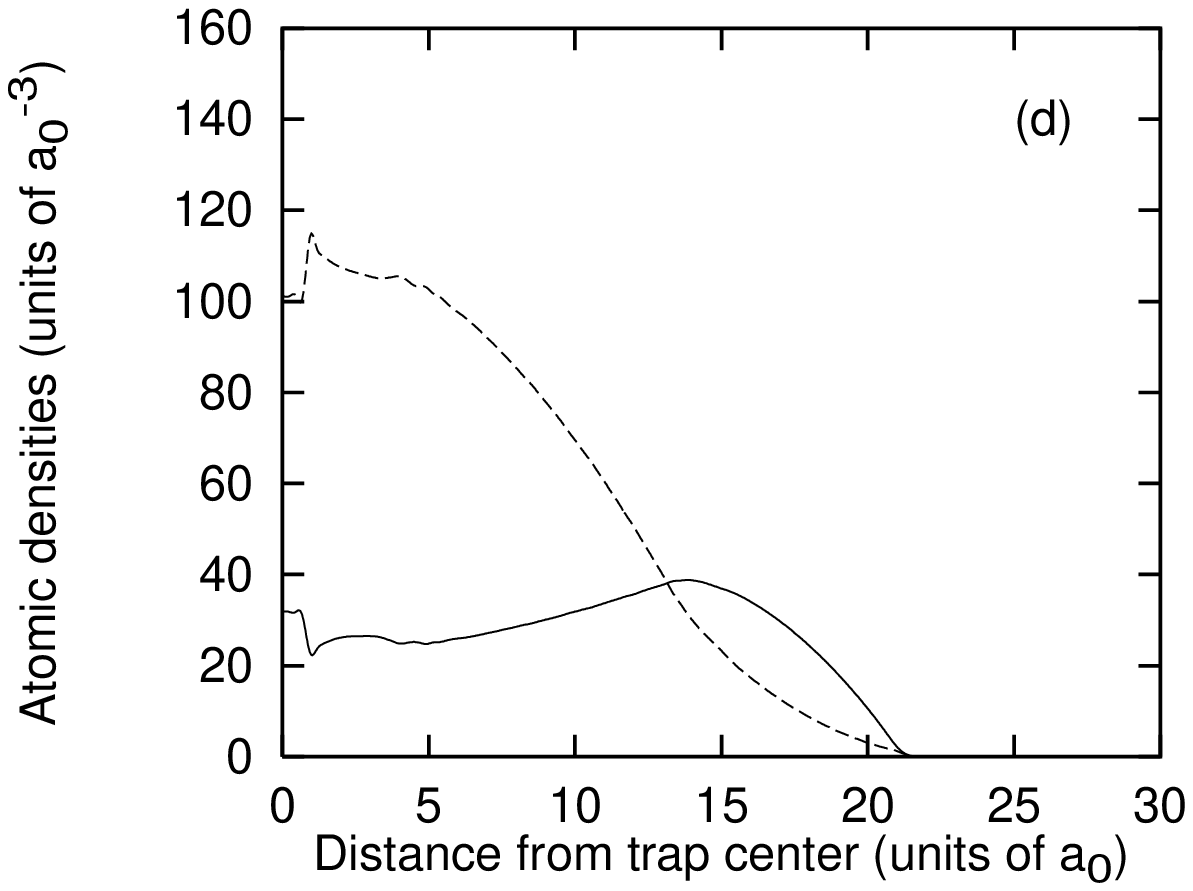,angle=270,width=5.0cm} &
\epsfig{file=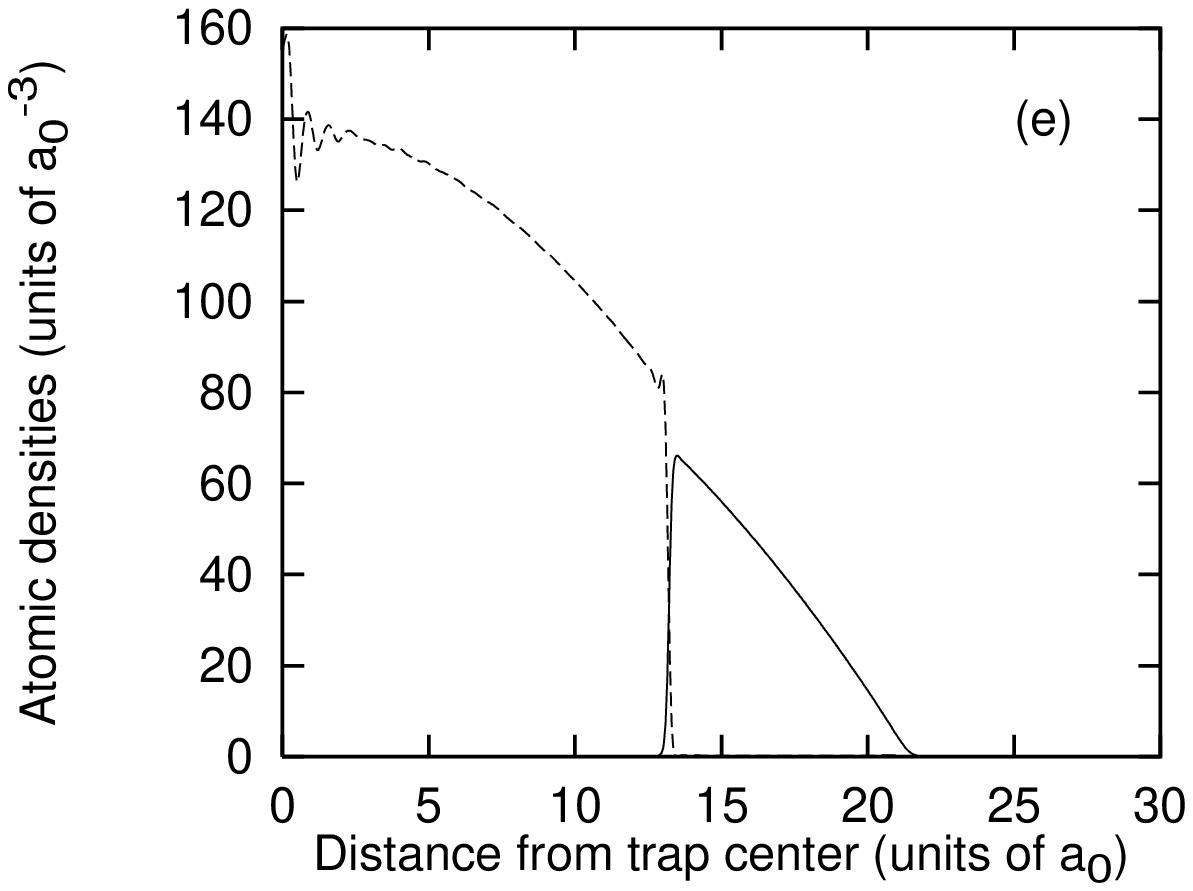,angle=270,width=5.0cm} &
\epsfig{file=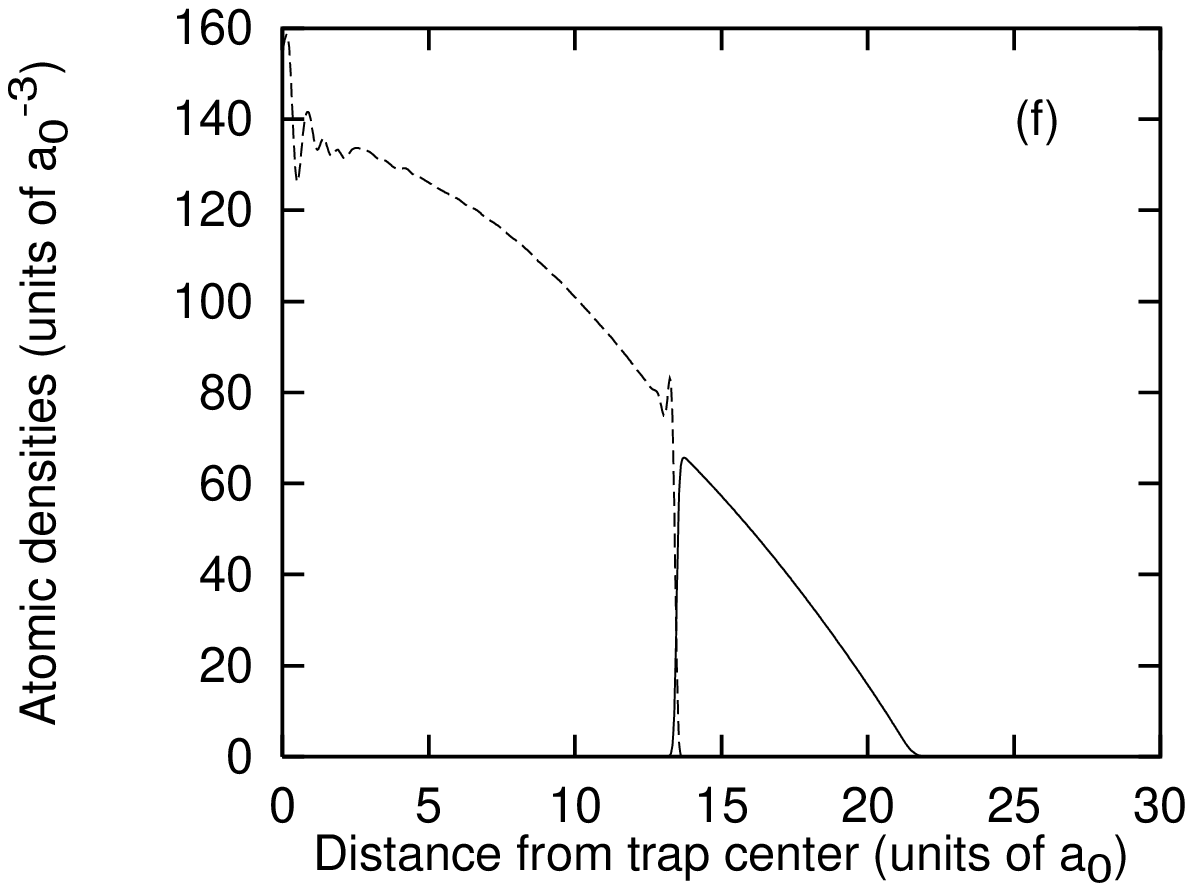,angle=270,width=5.0cm} \\
\end{array} $
\vspace{5 mm}
\caption{Spatial distribution of $10^6$ bosons (solid curves)
and $10^6$ fermions (dashed curves)
as function of distance from the trap center. 
In Figs. 2a-2f, $g=2.11 \hbar{\omega}a_0^3$, and the boson-fermion
coupling takes the values, $h=0,\ g/4,\ g/2,\ 3g/4,\
g,\ 5g/4$.}
\label{densplot}
\end{figure}

\begin{figure}
\centerline{
\epsfig{file=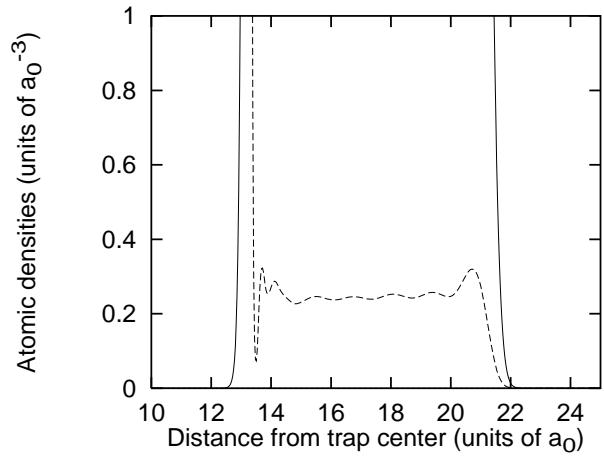,angle=270,width=8.5cm}}
\vspace{5 mm}
\caption{Spatial distribution of $10^6$ bosons (solid curves)
and $10^6$ fermions (dashed curves)
as function of distance to the trap center. 
$h=g=2.11 \hbar{\omega}a_0^3$ }
\label{magnify}
\end{figure}

We now turn to the case of equal numbers of bosons and fermions. The influence
of the inter-species interaction grows as the number of fermions is increased
with dramatical effects on the atomic distibutions, as we shall demonstrate.
We study the case of $10^6$ fermions, and the same number of bosons with an 
interaction strength of $g=2.11 \hbar\omega{a_0}^3$. We again expect that
for certain critical parameters, the components find it energetically 
favorable to separate into two distinct phases, but this time
bosons are expelled from the trap center, minimizing their
internal interaction energy by spreading in a 'shell' around a
fermionic bubble. Figs.~\ref{densplot} and~\ref{magnify} present our 
results. The essential features are again the spatial separation of the 
two components, this time manifesting itself by the exclusion of the 
bosonic component from the trap center and the existence of a constant fermion 
density through the boson distribution for $h=g$. For a different 
choise of parameters, for example by letting the fermions be trapped by a weaker 
potential, we are also capable of producing a multi-layered structure with  
fermions residing on both sides of the bosons.  

We notice that as the bosons are expelled from the center of the trap,
forming a 'mantle' around the fermions, the fermionic
component is compressed, having a higher peak density 
and covering a smaller portion of the trapping volume. A similar 
behavior has been noted for bi-condensate systems~\cite{sten2,bigelow3}.
One of the essential features predicted in the Thomas-Fermi approximation
is the existence of a 'plateau' of constant fermion density through
the boson distribution for $h=g$. As illustrated by Fig.~\ref{magnify},
which is just a magnification of the central parts of 
Fig.~\ref{densplot}e, this phenomenon also appear in our quantum mechanical
treatment, although with the parameters chosen it does not involve quite 
as many particles as obtained from the semi-classical calculations in~\cite{klaus1}. 

\begin{figure}
\centerline{
\epsfig{file=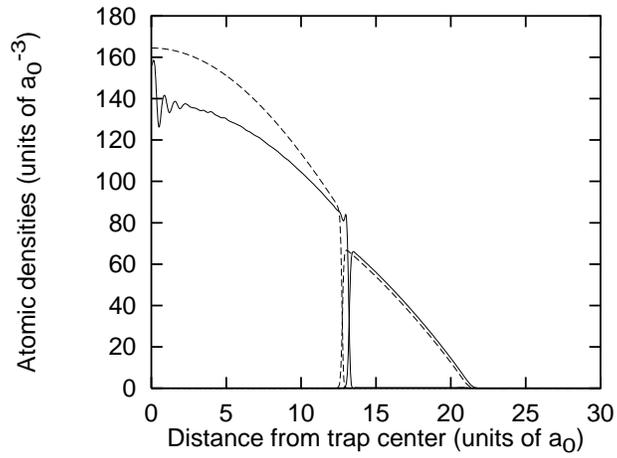,angle=270,width=8.5cm}}
\vspace{5 mm}
\caption{Comparison of density profiles calculated by using the
Slater-determinant method (solid lines) and the Thomas-Fermi
approximation (dashed lines) for the fermion density. $N_B=N_F=10^6$, $h=g=2.11
\hbar{\omega}a_0^3$.}
\label{compare}
\end{figure}

It is interesting to compare the above mentioned results with those 
obtained by treating the fermions in the Thomas-Fermi approximation.
This is done in Fig.~\ref{compare} for $N_B\!=\!N_F\!=\!10^6$, 
$h\!=\!g\!=\!2.11 \hbar\omega a_0^3$,
and we note that the semi-classical description
gives a qualitatively correct description, in that it reliably
predicts the phase separation. Thus it is reasonable to use this
approximative treatment of the fermions in the anisotropic case,
where the exact description is too cumbersome. 

\subsection{Anisotropic trap}
We now turn our attention to the anisotropic potentials, where we will
use only the semi-classical Thomas-Fermi approximation for the fermion density. 
We aim to reveal similar variations in the ground state density
profiles as for the isotropic trap, but going to higher dimensions
we now have the opportunity to investigate the phenomenon of
spatial symmetry breaking.
Intuitively, we assume that for critical parameters it may be
preferable for the two components to break mirror symmetry ($z\rightarrow -z$),
thereby  minimizing their mutual interaction, especially in elongated 
traps. Such a behavior has been predicted by \"{O}hberg and Stenholm 
for bi-condensates in two dimensions~\cite{sten2}. 

\begin{figure}
\centerline{
\epsfig{file=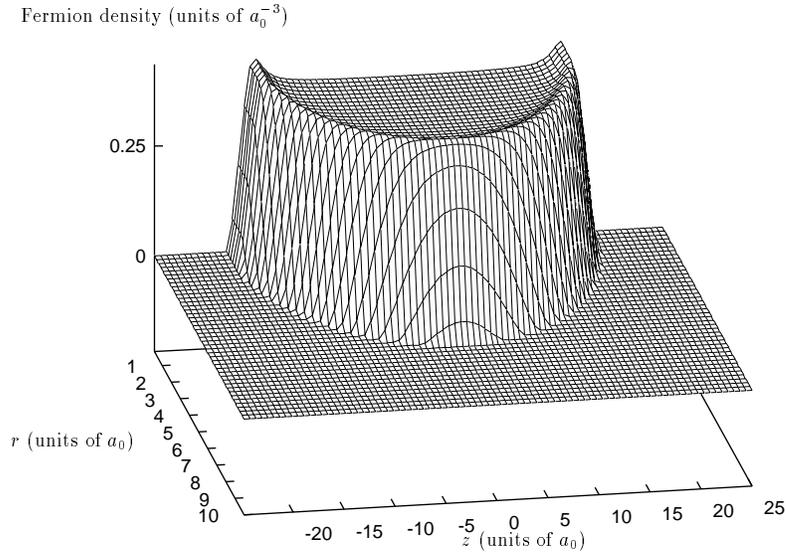,angle=270,width=15cm}}
\caption{Spatial distribution of 1000 fermions
in an anisotropic trap with $\lambda=1/\sqrt{8}$. 
$g=0.015 \hbar{\omega}a_0^3$, and the boson-fermion
coupling takes the values, $h=g$.
The Bose-Einstein condensate component of $10^6$ atoms 
is unaffected by the fermions and extends out to the distance $r\sim 7a_0$.}
\label{fermiplot2}
\end{figure}

\begin{figure}
\centerline{
\epsfig{file=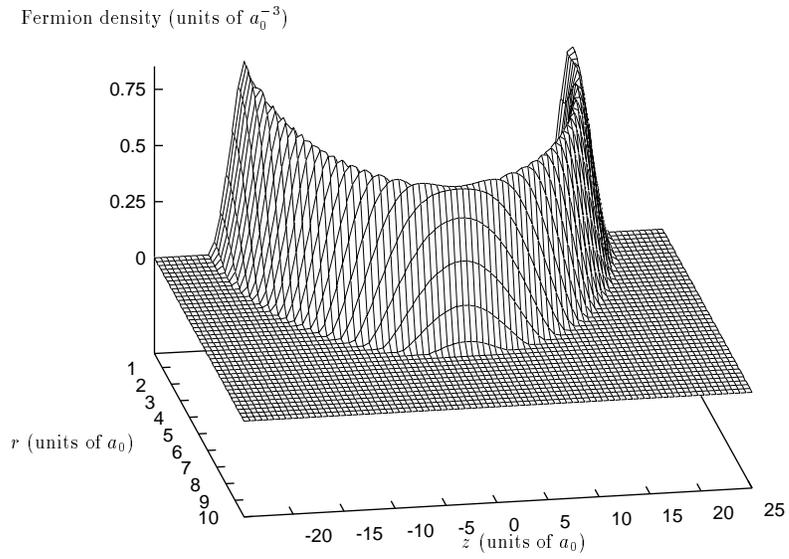,angle=270,width=15cm}}
\caption{Same as Fig.~\ref{fermiplot2}, but with $h=2g$.}
\label{fermiplot3}
\end{figure}

It remains to be demonstrated though, that the features described
in the case of the isotropic trap are still essential, when we 
consider the anisotropic scenario relevant in comparison with currently
experimentally feasible setups.
We present in Figs.~\ref{fermiplot2} and~\ref{fermiplot3} the analog of 
Fig.~\ref{fermiplot} with
the same choise of parameters and $\lambda=1/\sqrt{8}$, \emph{i.e.} the
inverse of the value for the current traps which have the strongest confinement 
along the $z$-axis. We notice the appearance 
of the same qualitative features as in the isotropic trap, that is component 
separation for $h>g$ and a plateau of constant fermion density for $h=g$.

The $10^6$ bosons are in the condensate which is unaffected in form and location
by the presence of the relatively few fermions.
Not shown in Figs.~\ref{fermiplot2} and~\ref{fermiplot3} is the distribution of 
fermions for $h$ smaller than $g$. In this case the fermionic component
overlaps the boson cloud at the center of the trap.

To address the issue of symmetry breaking we adopt the same 
procedure as \"{O}hberg and Stenholm~\cite{sten2}.
This offers only suggestive evidence that symmetry breaking may occur. 
To investigate this behavior correctly one must use an altogether 
different approach, minimizing the energy functional to find the ground-state density 
profile~\cite{gordon}. The point is that the solutions of the 
Gross-Pitaevskii equation are stationary points of the energy 
functional, not necessarily corresponding to minima. 
They may therefore be unstable in certain parameter regions.
It is possible though to single out the more stable of two
configurations by comparing their total energy as this is minimum in
equilibrium.

The total energy functional of the two-component system is a sum of 
four terms
\begin{equation}
E = T_B + T_F + V_{ext} + V_{int}.
\label{energyfunc}
\end{equation}
The first term is the boson kinetic energy
\begin{equation}
T_B = \int d^{3}r \, \frac{\hbar^2}{2M}|\nabla \psi(\vec r)|^2.
\label{boskin}
\end{equation}
As a fermion with wave number $\vec{k}(\vec r)$
has a kinetic energy of $\hbar^2k^2/2M$, the total fermionic
contribution to the kinetic energy is found by integrating this local term over all
of phase-space, weighted by the phase-space density, $1/(2\pi)^3$,
\begin{eqnarray}
T_F & = & \int \frac{d^{3}r}{(2\pi)^3} \, \int_{0}^{k_F(\vec r)} d^{3}k \, 
\frac{\hbar^2 k^2}{2M} \nonumber \\
 & = & \int \frac{d^{3}r}{2\pi^2} \, \frac{\hbar^2}{10M} \left[
6\pi^2 n_F(\vec r) \right]^{5/3}.
\label{fermikin}
\end{eqnarray} 
Calculating the potential energy terms is easy, as they involve
only integrals over the atomic densities
\begin{eqnarray}
V_{ext} & = & \int d^{3}r \, {1\over 2} M\omega^2 r^2 \left[ n_B(\vec r)+n_F(\vec r)
\right] \\
V_{int} & = & \int d^{3}r \, n_B(\vec r)\left[ g n_B(\vec r) +
hn_F(\vec r) \right].
\label{poten}
\end{eqnarray}
We have chosen the number of atoms to be $N_B=N_F=10^3$, while the asymmetry
parameter is still set to $\lambda=1/\sqrt{8}$. The interaction parameters are
$g=6.67\hbar{\omega}a_0^3$ and $h=5g$. 

\begin{figure}
\centerline{
\epsfig{file=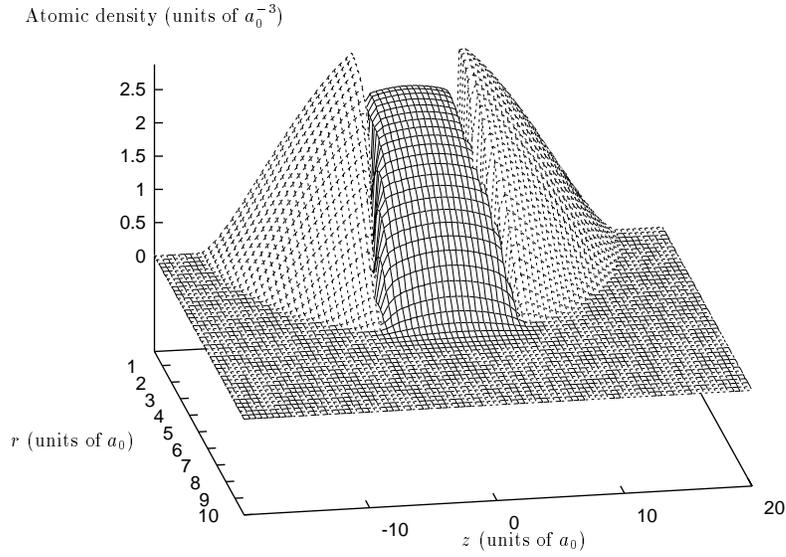,angle=270,width=15cm}}
\caption{The 'boson-burger' configuration of 1000 bosons (solid lines) and 1000 
fermions (dashed lines) in a prolate harmonic trap with $\lambda=1/\sqrt{8}$. 
The strength of the interparticle interactions are $g=6.67 \hbar{\omega}a_0^3$ and $h=5g$. 
The iteration was started with two Gaussian density profiles located oppposite 
to each other, away from the center.}
\label{boseburger}
\end{figure}

Starting the iteration with two well separated clouds displaced along the 
cylinder axis, \emph{i.e.} along the direction of the weaker trapping potential,
the calculation converges to a situation where the fermions
localize on both sides of a central concentration of the Bose condensate: a
'boson-burger', see Fig.~\ref{boseburger}.
Initiating the calculation with two overlapping clouds
in the center of the trap results in just the reversed situation: a 'fermion-burger',
consisting of a central fermionic part surrounded on two sides by bosons, but this
configuration has a larger energy. The 'boson-burger' seems to be the stable solution.
 
\begin{figure}
\centerline{
\epsfig{file=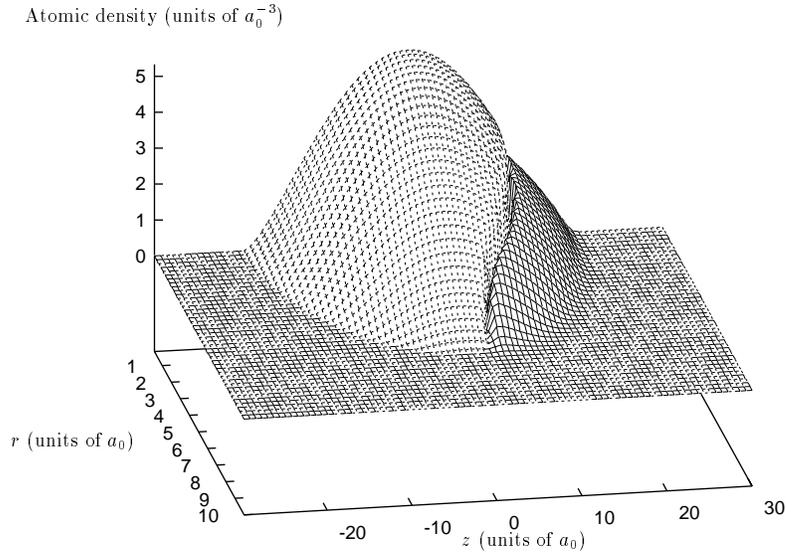,angle=270,width=15cm}}
\caption{Spatial density of 5000 fermions and 1000 bosons in a prolate trap
with $\lambda=1/\sqrt{8}$ and interaction strengths $g=6.67 \hbar{\omega}a_0^3$ and $h=5g$.
The bosons (solid lines) and the fermions (dashed lines) localize in separate regions
of space due to their strong repulsion and the elongated nature of the trap.}
\label{symbreak}
\end{figure}

In Fig.~\ref{symbreak} we show the spatial distribution of 5000 fermions and 1000
bosons. The particles feel the same trapping potential as in Fig.~\ref{boseburger}, 
and the interaction strengths are kept unchanged. 
This configuration is the result when the starting
point of the calculation is two separated clouds. When we start by placing both
species at the center of the trap we achieve again the 'fermion-burger', but
at a higher energy. Thus we conclude that in this region of parameter space
the system is unstable against breaking of the reflection symmetry.

We note that our approach provides two degenerate symmetry broken states, the 
one in Fig.~\ref{symbreak} and its mirror image in the $xy$-plane.
Going beyond our theoretical treatment (Hartree), we may construct
superpositions of these two macroscopically states which do not break the 
spatial symmetry. One of these states will have a lower energy, but such 
a 'Schr{\"o}dinger-cat' state is exceedingly
complicated to prepare, {\emph{c.f.}} the discussion in~\cite{klaus2}.
Thus the symmetry broken solution is most likely to be observed in an
experiment.   

\section{Conclusion}
In this paper we have investigated the zero temperature ground state
of a mixture of boson and fermion gases in both isotropic and
anisotropic trapping potentials. We have addressed the
issue of component separation using nummerical techniques to solve
coupled equations for the spatial density of the two species. Our 
calculations have confirmed and expanded upon the results of a
previous paper~\cite{klaus1}, which treated the problem only within
the Thomas-Fermi approximation for both components and which analyzed only
the case of an isotropic trap. We have confirmed the existence of three
distinct states of the system under variation of the ratio of the interaction 
strengths $h/g$: For small values of this parameter the gases are
interpenetrable, overlapping throughout the occupied volume of the trap,
as their mutual repulsion is not strong enough to cause separation. When
the coupling strength $h$ exceeds the strength of the boson-boson interaction
one of the species is expelled from the center of the trap. The spatial
configuration in this case depends on the symmetry of the trapping potential.
In an isotropic trap the separated phase is rotationally symmetric, the excluded
component constitutes a spherical shell wrapped around a centrally compressed 
bulk. The anisotropic trap however has a parameter region where a breaking of
symmetry ($z\rightarrow -z$) may occur, and we have demonstrated such forms.
In the limiting case $h=g$ there exists the possibility for the fermions to have
a constant spatial density where the bosons are localized.

An aspect of this work is the availability of an almost
isolated degenerate Fermi gas through the complete separation of the two species.
The trapped, degenerate Fermi gas is interesting in view of the possibility of a BCS
transition when two spin states are trapped simultaneously~\cite{burnett6,stoof1,stoof2}
and because of the analogies between this system and both atomic nuclei and the 
interior of neutron stars.

The details of sympathetically cooling the Fermi gas to the degeneracy level through
thermal contact with the Bose condensate are of course of great
importance in further research~\cite{timmer1}. In general the investigation
of the cooling ability of the condensate should not be restricted to fermionic
impurities. In view of the recent trapping of simple molecules in both optical
~\cite{stapelfeldt} and magnetic~\cite{cahtrap} potentials, also more complex
solutes with several internal degrees of freedom pose an interesting challenge
for future research.

Another direction worth noticing is the prospect of trapping a boson-fermion mixture  
in the periodic potential of an optical lattice~\cite{klaus3}, both in
its own right and as a study of solid state phenomena. With quantum gases
well beyond the degeneracy level a complete filling of the potential wells
may well be expected~\cite{zoller2}.

Finally it should be mentioned that in this work we have concentrated solely on systems 
with a positive coupling strength $h$. Allowing the interaction between the species to
become attractive is known to induce a dramatic change in the macroscopic behavior of the
system as it becomes unstable against collapse for large negative values of 
$h$~\cite{klaus1}. We are currently setting up calculations to investigate this phenomenon 
in detail using the numerical procedure developed in this work.
 

\newpage

\newpage

\newpage

\newpage

\newpage

\newpage

\newpage
 
\end{document}